\documentclass[11pt]{article}

\usepackage[margin=1.1in]{geometry}

\usepackage{tabto}
\usepackage{amsmath}
\usepackage{boldline}
\usepackage[T1]{fontenc}
\usepackage{graphicx}
\newtheorem{definition}{Definition}[section]
\newtheorem{problem}{Problem}[section]

\begin{document}
\title{Multiple Double Arithmetic on NVIDIA Tensor Cores}
%
%

\author{
Howard Chen\thanks{University of Illinois at Chicago,
Department of Mathematics, Statistics, and Computer Science,
851 S. Morgan St. (m/c 249), Chicago, IL 60607-7045,
Email: {\tt hchen221@uic.edu}.}
\and
Jan Verschelde\thanks{University of Illinois at Chicago,
Department of Mathematics, Statistics, and Computer Science,
851 S. Morgan St. (m/c 249), Chicago, IL 60607-7045,
Email: {\tt janv@uic.edu}, URL: {\tt http://www.math.uic.edu/$\sim$jan}.}
}
\date{1 July 2026}

\maketitle              
\begin{abstract}
A multiple double is an unevaluated sum of doubles.
An NVIDIA tensor core is a specialized high performance
compute core for matrix multiplication.
The Ampere A100, released in 2020, introduced tensor cores
capable of 64-bit floating-point arithmetic.
Every multiple double arithmetical operation requires renormalization,
which involves branching, for which tensor cores are unsuited.

To solve this problem caused by renormalization,
we apply a solution similar to the Ozaki scheme
[Ozaki et al, Numerical Algorithms, 2012].
Our software is available under the GPU GPL license on github.
\end{abstract}
\section{Introduction}

An algorithm is {\em robust} if it does not fail for small perturbations
of degenerate inputs.  With multiple double arithmetic we can obtain
more accuracy than what can be provided with 64-bit doubles.

\begin{definition}[multiple double]
A {\em multiple double} is an unevaluated sum of nonoverlapping doubles.
\end{definition}
The algorithms for multiple double arithmetic originated in the
late sixties, in efforts~\cite{Dek71} to extend the accuracy of the then
commonly used 32-bit floating-point arithmetic.
In~\cite{HLB01} and~\cite{She97}, algorithms are defined
for quad double arithmetic, and also described in~\cite{MBDJJLMRT18}.
QDlib~\cite{HLB01} and CAMPARY~\cite{JMPT16} are two software packages
for quad double and general multiple double arithmetic respectively.

If the result of a computation can be represented exactly by
a 64-bit double, then the next double in the sequence of multiple
doubles represents the working precision in which the computation
was executed.  
This property justifies the term error-free transformations~\cite{Rum10}
when referring to multiple double arithmetic.
The drawback of working with multiple double arithmetic is its
cost overhead.  For instance, in order to compensate for the cost
of quad double arithmetic, one needs to run computations with
teraflop performance.

In order to offset the cost of multiple double algorithm,
we examine the application of Graphics Processing Units (GPUs),
and in particular in this paper, NVIDIA Tensor Cores.

\begin{definition}[NVIDIA Tensor Core]
An {\em NVIDIA Tensor Core} is a specialized high performance
compute core for matrix multiplication.
\end{definition}
Introduced in 2020, the NVIDIA ``Ampere'' A100 Tensor Cores
offer IEEE-compliant 64-bit floating-point (FP64) tensor core 
instructions~\cite{NVIDIA_A100}.
An appealing aspect of tensor cores are their increased performance,
relative to the regular cores.  In particular,
the theoretical peak performance of the FP64 (non-tensor) cores 
is 9.7 TFLOPS (teraflops), whereas 19.5 TFLOPS is the theoretical peak 
performance of the FP64 tensor cores.

Then our question becomes:
{\em Are tensor cores useful for multiple double arithmetic?}
At first sight, the answer is simply no because of the following.
\begin{problem} [renormalization]
Multiple double arithmetic requires a renormalization after
every arithmetical operation.  
This renormalization involves branching and tensor cores are
specialized for matrix multiplication.
\end{problem}

Our solution is similar to the application of 
the Ozaki scheme~\cite{OOOR12},
used in mixed precision computations~\cite{AABetal21}, 
\cite{HM22}, \cite{KLBRMSW26}, \cite{MOOI20}.
In mixed precision, accurate results are computed via lower
precision arithmetic.  Multiple double arithmetic is similar,
using double precision arithmetic to multiple the accuracy
of the results.  Double precision arithmetic has the benefit
of a wider range of exponents compared to what is available
on lower precision processors.

New algorithms for extended precision floating-point arithmetic
that avoid branching were proposed in~\cite{ZA25},
available in a registered Julia package~\cite{Zha26}.
The vectorization of multiple double arithmetic is achieved
via branch-free algorithms, presented in~\cite{Kou26} for double double,
triple double, and quad double addition and multiplication,
with comparisons made to MPLAPACK~\cite{Mah22}.
While related to our work, those branch-free algorithms are not suitable
for execution by tensor cores.


The next section introduces the rewriting of a product of two
double double matrices as a product of two double matrices,
followed by a description of balancing algorithms to ensure
double double accuracy.  
The algorithms work for general multiple double arithmetic.
Section~4 presents our software and experimental results.  

\section{Rewriting Products of Double Double Matrices}

The fraction {\tt f} of a double $x$ has 52 bits,
and we can distribute the bits into 4 parts 
{\tt f0}, {\tt f1}, {\tt f2}, {\tt f3}, each of 13 bits:

\begin{center}
{\small
\begin{verbatim}
 f : 1001 0001 1010 0110 0110 0100 1110 1101 0010 0011 0100 1100 1100 
f0 : 1001 0001 1010 0000 0000 0000 0000 0000 0000 0000 0000 0000 0000 
f1 : 0000 0000 0000 0110 0110 0100 1100 0000 0000 0000 0000 0000 0000 
f2 : 0000 0000 0000 0000 0000 0000 0010 1101 0010 0010 0000 0000 0000 
f3 : 0000 0000 0000 0000 0000 0000 0000 0000 0000 0001 0100 1100 1100 
\end{verbatim}
}
\end{center}

Then we compute with 4 doubles {\tt x0}, {\tt x1}, {\tt x2}, {\tt x3},
we call them {\em quarters},
with exponents {\tt 0}, {\tt -13}, {\tt -26}, {\tt -39}.

\begin{center}
\begin{verbatim}
x0 : 1 0010 0011 0100 000000000000000000000000000000000000000    0 
x1 : 1 1001 1001 0011 000000000000000000000000000000000000000  -13
x2 : 1 0110 1001 0001 000000000000000000000000000000000000000  -26
x3 : 1 0100 1100 1100 000000000000000000000000000000000000000  -39
\end{verbatim}
\end{center}
The zero bits at the end allow for error free computations:
\begin{itemize}
\item Multiplying two quarters leaves at least 26 zero bits at the end.
\item Long sequences of inner products can be made 
      {\em without losing any bits.}
\end{itemize}

If we focus on the four quarters of the leading doubles:
\begin{equation}
a_{i,k} = (a_{i,k,0}, a_{i,k,1}, a_{i,k,2}, a_{i,k,3}), \quad
b_{k,j} = (b_{k,j,0}, b_{k,j,1}, b_{k,j,2}, b_{k,j,3}),
\end{equation}
then we can map the elements of matrices $A_{i,k}$ and $B_{j,k}$ as

\begin{equation}
   A_{i,k} =
   \left[
     \begin{array}{c}
       a_{i,k,0} \\
       a_{i,k,1} \\
       a_{i,k,2} \\
       a_{i,k,3} 
     \end{array}
   \right]^T, \quad
   B_{k,j} =
   \left[
     \begin{array}{cccc}
       b_{k,j,0} & b_{k,j,1} & b_{k,j,2} & b_{k,j,3} \\
           0     & b_{k,j,0} & b_{k,j,1} & b_{k,j,2} \\
           0     &     0     & b_{k,j,0} & b_{k,j,1} \\
           0     &     0     &     0     & b_{k,j,0} \\
     \end{array}
   \right],
\end{equation}
contributing to the product
\begin{equation} \label{eqproductconvolutions}
  C_{i,j} = 
  \left[
    \begin{array}{l}
       a_{i,k,0} b_{k,j,0} \\
       a_{i,k,1} b_{k,j,0} + a_{i,k,0} b_{k,j,1} \\
       a_{i,k,2} b_{k,j,0} + a_{i,k,1} b_{k,j,1} + a_{i,k,0} b_{k,j,2} \\
       a_{i,k,3} b_{k,j,0} + a_{i,k,2} b_{k,j,1} + a_{i,k,1} b_{k,j,2} 
     + a_{i,k,0} b_{k,j,3} 
    \end{array}
  \right].
\end{equation}
In~(\ref{eqproductconvolutions}), observe the convolutions
with the sum of the last two indices in the product of two numbers
equal to the index of the quarter.

A double double matrix, represented by two matrices 
$(A_\text{hi},A_\text{lo})$ of high and low doubles,
is mapped into a sequence of 8 matrices 
$(A_0$, $A_1$, $A_2$, $A_3$, $A_4$, $A_5$, $A_6$, $A_7)$,
where the first four matrices are the quarters of the high doubles,
and the last four matrices collects the quarters of the low doubles.
Similarly for the second double double matrix $(B_\text{hi},B_\text{lo})$,
the quarters are mapped into a sequence of 8 matrices
$(B_0$, $B_1$, $B_2$, $B_3$, $B_4$, $B_5$, $B_6$, $B_7)$.
Then the product of two double double matrices is rewritten
as the product of the matrices
\begin{equation} \label{eqrewrite}
   \overline{A} :=
   \begin{bmatrix}
      A_0 & A_1 & \cdots &A_7
   \end{bmatrix}
   \quad \mbox{and} \quad
   \overline{B} := 
   \begin{bmatrix}
       B_0 & B_1 & \cdots & B_7 \\
           & B_0 & \cdots & B_6 \\
           &     & \ddots & \vdots \\
           &     &        & B_0
   \end{bmatrix}
\end{equation}
which then leads to the product
$\overline{C}:=\begin{bmatrix}C_1&C_2&\dots&C_8\end{bmatrix}$,
computed as $\overline{A}\cdot\overline{B}$ by the tensor cores
in double precision.

Note that upon quartering, the size of the data increases by 4 times. 
For example, a quartered double double $n\times n$ matrix has 8 times 
the size of a regular $n\times n$ matrix of doubles, 
and a quartered quad double matrix has 16 times the size.
Instead of stacking quartered matrices into $\overline{A}$
and convoluting quartered matrices into $\overline{B}$,
the first matrix could be obtained from convoluting quartered matrices
and the second matrix could be a stacking of quartered matrices.

In the overview of the algorithm to multiply two double double matrices 
on tensor cores, we list the following steps:
\begin{enumerate}
\item Split all double doubles in 8 parts.
\item Rewrite into one single matrix product of double matrices.
\item Run the matrix multiplication on the tensor cores.
\item Extract the parts from the product.
\item Add the parts of the product into a double double matrix.
\end{enumerate}

The last step must be done in double double arithmetic, summing 
8 doubles into one double double for each element of the product.
This double double summation is massively parallel and 
can be executed on regular cores.

The main question then remains: {\em Is the result accurate?}

We call a sequence of positive doubles {\em balanced} if their
exponents lie on an equally spaced grid, for example: $0$, $-13$,
$-26$, $-29$, etc.  The inner product of two balanced sequences
then reduces to a sum of doubles with fractions that have only their 
first 26 bits nonzero, or equivalently, with 26 trailing zero bits.
Sums of doubles with 26 nonzero bit fractions are computed exactly
as long as there is no overflow, that is: if the sum fits in the 
52-bit size fraction of a double.
If $N$ is the size of such sum, then how large can $N$ be before
overflow occurs?  Consider:
\begin{equation} \label{eqNbound}
   N \underbrace{\left( \sum_{i=0}^{25} 2^i \right)}_{B} \leq 2^{52},
\end{equation}
where $B$ is the largest number with 26 leading nonzero bits.
Evaluating the inequality~(\ref{eqNbound}) yields 67,108,865
as the upper bound for~$N$.

The next section addresses the case when the doubles are not balanced.

\section{Balancing Algorithms}

We have to group the numbers by sign and size,
keeping in mind the loss of accuracy in floating-point arithmetic.

\begin{problem}
Subtracting floating-point numbers of the same magnitude
may lead to losing many significant bits in the fraction.
\end{problem}
Therefore, group elements by sign as follows:
\begin{itemize}
\item let $A_+$ contain all positive numbers of $A$, and
\item let $A_{-}$ contain all negative numbers of $A$, 
\end{itemize}
so $A = A_+ + A_{-}$, and likewise $B = B_+ + B_{-}$, then
\begin{equation}
  A B = (A_+ + A_{-}) (B_+ + B_{-})
      = A_+ B_+ + A_+ B_{-} + A_{-} B_+ + A_{-} B_{-}.
\end{equation}

\begin{problem}
Adding numbers with exponents of different magnitude 
may lead to losing many significant bits in the fraction.
\end{problem}
To solve this problem, write $A$ as $A_0 + A_1 + \cdots + A_{12}$,
where the index is the exponent of the leading quarter of the double,
corresponding to the 13 possible exponents of the leading quarters
in the distribution of the doubles of~$A$.
Do likewise for $B$ and execute many matrix multiplications.

\begin{problem}
A double double $x$ is represented by 
a high double $x_h$, and
a low double $x_\ell$.
The problem is then that $x_h$ and $x_\ell$ may have opposite signs.
\end{problem}
To solve this problem, assuming $x_h > 0$ 
and the exponent of $x_h$ is zero, do:
\begin{equation}
   b := 1 \times 2^{-52}; \quad
   x_h := x_h - b; \quad
   x_\ell := x_\ell + b.
\end{equation}
Obviously, the value $x_h + x_\ell$ of the double double does not
change by these steps.  Observe the following.
As the original $x_\ell < 0$, and in particular, 
as $x_\ell < -1 \times 2^{-52}$, we have $x_\ell + b > 0$,
and thus the new $x_\ell > 0$.
As $x_\ell$ grows in size, we may loose the last bit of $x_\ell$.
Therefore, in a multiple double, the last bit should then be added to the
next double in the sequence of doubles representing the multiple double,
using multiple double arithmetic.

\begin{problem}
If we quarter a double $x$ in four parts: $x_0$, $x_1$, $x_2$, and $x_3$,
we expect the exponents to be 0, $-13$, $-26$, and $-39$, respectively.
The problem is then that $x_1$, $x_2$, and $x_3$ may have many leading
zeros in their fractions.
Therefore, their exponents may be much less than $-13$, $-26$, and $-39$,
in the worst case: $-25$, $-38$, and $-51$,
causing great loss of accuracy.
\end{problem}
To solve this problem, 
if the exponents of $x_0$ and $x_1$ differ by more than 13, 
assuming $x_0 > 0$, and assuming $x_0$ belongs to $A_0$, 
i.e.: its exponent is zero, do: 
\begin{equation}
   b := 1 \times 2^{-14}; \quad
   x_0 := x_0 - b; \quad
   x_1 := x_1 + b.
\end{equation}
Again obviously, the value of $x_0 + x_1$ remains invariant by
these steps.  Observe the following.
Subtracting $b$ does not alter the exponent of $x_0$.
As the exponent of $x_1$ is less than $-13$, $x_1 + b$ has exponent $-13$,
and thus the exponents of $x_0$ and $x_1$ are 0 and $-13$ respectively.
To ensure that $x_1$ still has many trailing zero bits,
the bits in the fraction after the 13-th position have to be added
to the next quarter $x_2$, which may be beneficial if the exponent
of $x_2$ is less than $-26$.
Similar steps and arguments apply then to adjust $x_2$ and $x_3$
so they have the desired exponents of $-26$ and $-39$ respectively.

A special case occurs when one double in the sequence representing
a multiple double is equal to zero.  In that case, we assume the
unnormalized representation of zero with the desired exponent,
and just do nothing.  In any case, when the exponents of the doubles
lie on a grid, the argument explained at the end of the previous 
section applies and very large inner products are computed accurately.

\section{Multiple Double Matrix Multiplication}

In our software, we applied a top down and a bottom up methodology.
The bottom up method does not rely on any available code for
matrix multiplication on tensor cores, whereas the top down method
started from adjusting the {\tt dmmaTensorCoreGemm} from the
CUDA samples~\cite{NVIDIA_samples}.

\subsection{Hardware and Software}

We used the CUDA Software Development Kit, {\tt nvcc} version 12.4,
on two different NVIDIA GPUs, in computers with specifications listed below:

\begin{enumerate}
\item Microway GPU Server with NVIDIA Ampere A100 80GB GPU,
      with two Intel Xeon 5318Y 2.1 GHz 24-core CPUs and 256~GB
      internal memory, running Rocky Linux 9.3.
      The host code is compiled with {\tt g++} 11.4.1.
      Using shared memory, our modified {\tt dmmaTensorCoreGemm}
      reached 13.75 FP64 TFLOPS.
\item NVIDIA RTX 4080 housed in 2023 Alienware gaming laptop,
      with a 24-core Intel i9-13900~HX at 2.2GHz and 32~GB
      internal memory, running Windows 11 and the
      2022 Community Visual Studio version.
      The FP64 Tensor Cores on RTX 4080 gave a performance
      of 0.50 TFLOPS when running the modified {\tt dmmaTensorCoreGemm}.
\end{enumerate}

\noindent Our software is available under the GPU GPL license, at
{\tt https://github.com/}
\begin{itemize}
\item {\tt janverschelde/PHCpack/src/GPU/TensorCores}
\item {\tt hchen221/multiple-double-arithmetic-matrix-matrix-multiplication}
\end{itemize}

The PHCpack source code~\cite{Ver99}
contains the codes for the multiple double arithmetic, 
taken from QDlib~\cite{HLB01} for double doubles and quad doubles
and generated with CAMPARY~\cite{JMPT16} for octo doubles and hexa doubles,
applied in~\cite{Ver21}, \cite{Ver22}, and~\cite{Ver24}.
The first experiments with vectorized multiple double arithmetic
in PHCpack were reported in~\cite{Ver25}.
The {\tt TensorCores} folder implements the top down method,
where as the other software applies the bottom up method.


In all experiments, random numbers were generated.

\subsection{Bottom Up Method}

Matrix products were computed using 2 methods. 
For the first method, using {\tt dmmaTensorCoreGemm} as a reference,
we defined a WMMA kernel\footnote{WMMA abbreviates Warp Matrix Multiply
Accumulate.} for a matrix matrix multiplication 
which was executed on $\overline{C}=\overline{A}\cdot\overline{B}$, 
as defined in~(\ref{eqrewrite}), then extracted the parts from $\overline{C}$ 
and merged them into a matrix of double doubles~$C$.
For the second method, we computed the matrix product $A\cdot B$ of 
double doubles, where a single thread executes an inner product computation.
Addition and multiplication operations within the inner product computations
were done using double double arithmetic.

Denote $C_\text{TC}$ as the result from the Tensor Core computation 
and $C_\text{CUDA}$ as the result from the regular CUDA core computation. 
Let the maximum error be 
\begin{equation}
   \epsilon_{\max} 
   := \max_{i,j\in[n]^2} 
   \left| \vphantom{\frac{1}{2}} C_\text{TC}[i,j]-C_\text{CUDA}[i,j] \right|
\end{equation}
We also tracked the amount of time it takes each method for different 
values of~$n$.  
The time for the Tensor Cores $t_\text{TC}$ was assessed as the
amount of time it takes to construct $\overline{A}$ and $\overline{B}$,
execute the Tensor Core kernel (a.k.a. WMMA kernel), and to assemble
the result into multiple doubles. 
The time for the CUDA kernel $t_\text{CUDA}$ was assessed as the amount 
of time it takes to execute a the CUDA kernel alone. 
For the sake of a direct comparison of matrix multiplication execution, 
we also tracked the time to execute the Tensor Core kernel 
alone $t_\text{WMMA}$.
Results from sample runs for double doubles 
and quad doubles are listed in Table~\ref{tabtimes1}.

\begin{table}[hbt]
\begin{center}
\caption{Units of all times $t_{\text{WMMA}}$, $t_{\text{TC}}$,
$t_{\text{CUDA}}$ are seconds:
$t_{\text{WMMA}}$ measures the execution of a single WMMA kernel,
while $t_\text{TC}$ is the time to make $\overline{A}$ and $\overline{B}$,
execute the WMMA kernel, and combine the result into a multiple double;
$t_\text{CUDA}$ is the time of a multiple double matrix multiplication
with a regular CUDA kernel.
Corresponding to the times are the flops
$f_\text{WMMA}$ and $f_\text{CUDA}$, in teraflops units.}
\label{tabtimes1}
\begin{tabular}{|c V{4} c|c|c|c V{4} c|c|} \hline
    & \multicolumn{4}{c V{4}}{double double} 
    & \multicolumn{2}{c|}{quad double} \\ \cline{2-7}
$n$ & 1024 & 2048 & 3072 & 4096     & 1024     & 2048 \\ \hlineB{4}
$t_\text{WMMA}$   & 0.000108 & 0.000106 & 0.000106 & 0.000257
                  & 0.000118  & 0.00011 \\\hline
$t_\text{TC}$     & 0.379792 & 1.19952 & 3.16265 & 7.93621
                  & 0.752966 & 4.23463 \\ \hline
$t_\text{CUDA}$   & 0.096794 & 0.450174 & 1.19599 & 2.33072
                  & 0.285909 & 1.57585 \\ \hlineB{4}
$f_\text{WMMA}$   & \multicolumn{1}{c|}{4.6} 
                  & \multicolumn{1}{c|}{2.7} 
                  & \multicolumn{1}{c|}{2.4} 
                  & \multicolumn{1}{c V{4}}{1.7} 
                  & \multicolumn{1}{c|}{3.6} 
                  & \multicolumn{1}{c|}{1.8} 
\\ \hline
$f_\text{CUDA}$   & \multicolumn{1}{c|}{2.8} 
                  & \multicolumn{1}{c|}{2.9} 
                  & \multicolumn{1}{c|}{2.8} 
                  & \multicolumn{1}{c V{4}}{2.9} 
                  & \multicolumn{1}{c|}{3.7} 
                  & \multicolumn{1}{c|}{3.9} 
 \\ \hlineB{4}
$\epsilon_{\max}$ & ~~$3.5\mbox{E}{-28}$~~ & ~~$8.7\mbox{E}{-28}$~~
                  & ~~$1.6\mbox{E}{-27}$~~ & ~~$2.4\mbox{E}{-27}$~~ 
                  & ~~$2.0\mbox{E}{-59}$~~ & ~~$4.1\mbox{E}{-59}$~~
        
                  \\ \hline
\end{tabular}
\end{center}
\end{table}

The execution of the WMMA kernel alone ($t_\text{WMMA}$) is significantly 
shorter than that of a regular CUDA core ($t_\text{CUDA}$), 
but the full process involving the usage of Tensor Cores ($t_\text{TC}$)
takes significantly longer than the CUDA cores. 
The extent to which $t_\text{TC}$ increases with respect 
to $n$ is greater than that of $t_\text{CUDA}$,
due to the size of $\overline{A},\overline{B}$.
The arithmetic intensity (\#operations per data unit) 
of multiple double arithmetic benefits regular CUDA kernels,
whereas matrix multiplication in double precision is memory bound,
rather than compute bound.

\subsection{Top Down Method}

The top down method applies the {\tt dmmaTensorCoreGemm} of the
CUDA software development kit on the product of double double matrices,
rewritten as a product of double matrices 
which contain the parts of the doubles of the input matrices.
In the top down method, the dimensions of the {\tt dmmaTensorCoreGemm} are
8192-by-4096 for the first matrix and 4096-by-8192 for the second matrix.
Table~\ref{tabDMMAdims} lists the dimensions adapted for higher precisions.

\begin{table}[hbt]
\begin{center}
\caption{Dimensions of the matrix $A$ of {\tt dmmaTensorCoreGemm}
adapted for double double (dd), quad double (qd), octo double (od),
and hexa double (hd).}
\label{tabDMMAdims}
\begin{tabular}{r V{4} r|r|r|r|r}
 & \multicolumn{1}{c|}{d}
 & \multicolumn{1}{c|}{dd}
 & \multicolumn{1}{c|}{qd}
 & \multicolumn{1}{c|}{od}
 & \multicolumn{1}{c}{hd} \\ \hlineB{4}
   \#rows of $A$ & 8192 & 1024 & 512 & 256 & 128 \\
\#columns of $A$ & 4096 &  512 & 256 & 128 &  64
\end{tabular}
\end{center}
\end{table}

A 1024-by-512 double double matrix is rewritten into
the first 8192-by-4096 matrix of double numbers
in the input of {\tt dmmaTensorCoreGemm}.  
The second 4096-by-8192 input matrix is obtained by
rewriting a 512-by-8192 double double matrix.
Each time the precision level doubles, the dimensions of the
multiple double matrices are cut in half,
as illustrated by Table~\ref{tabDMMAdims}.


With the version of {\tt dmmaTensorCoreGemm} which uses shared
memory, a performance of 13.75 TFLOPS is achieved,
accomplishing double double accuracy.
For a comparison: 13.75 > 9.7;
the theoretical peak performance of the regular CUDA cores
on the NVIDIA A100 GPU is 9.7 TFLOPS with FP64 arithmetic.

\section{Conclusions}


To apply FP64 tensor cores to multiply multiple double matrices,
the doubles in the matrices are partitioned introducing trailing 
zeros in the fractions of the doubles.  
Grouping matrices by signs and exponents, balancing algorithms 
redistribute bits in the parts of the fractions
to ensure the exponents of the parts remain on grid.
Our free and open source software demonstrates the acceleration
of multiple double arithmetic with FP64 tensor cores.


Future work will apply this multiple double tensor core matrix multiplication
to accelerate the blocked Householder QR, extending~\cite{Ver22}, 
with the application to compute Taylor series~\cite{Ver24}.

\bibliographystyle{plain}

\end{document}